\documentclass[fleqn,usenatbib]{mnras}

\usepackage[T1]{fontenc}
\usepackage{ae,aecompl}
\usepackage{threeparttable}
\usepackage{ulem}

\usepackage{graphicx}	
\usepackage{amsmath}	
\usepackage{amssymb}	
\usepackage{multirow}

\title[Interstellar extinction correction using HeI lines]{Interstellar extinction correction in ionised regions using HeI lines}

\author[S. Zamora et al.]{
S. Zamora,$^{1, 2}$\thanks{E-mail: sandra.zamora@uam.es}\thanks{PhD fellow of Ministerio de Educación y Ciencia, Spain}
Ángeles I. Díaz,$^{1, 2}$, Elena Terlevich$^3$\thanks{Visiting Professor, UAM, Madrid}\thanks{Visiting Astronomer, IoA, Cambridge, UK}, Vital Fernández$^{4,5}$
\\
$^{1}$Departamento de Física Teórica, Universidad Autónoma de Madrid, 28049 Madrid, Spain\\
$^{2}$CIAFF, Universidad Autónoma de Madrid, 28049 Madrid, Spain\\
$^{3}$Instituto Nacional de Astrofísica, Óptica y Electrónica (INAOE), Mexico\\
$^{4}$Departamento de Astronom\'ia, Universidad de La Serena, Av. Juan Cisternas 1200 Norte, La Serena, Chile\\
$^{5}$Instituto de Investigaci\'on Multidisciplinar en Ciencia y Tecnolog\'ia, Universidad de La Serena, Ra\'ul Bitr\'an 1305, La Serena, Chile\\
}

\date{Accepted 2022 July 28. Received 2022 July 22; in original form 2022 February 7}

\pubyear{2020}

\begin{document}
\label{firstpage}
\pagerange{\pageref{firstpage}--\pageref{lastpage}}
\maketitle

\begin{abstract}
The logarithmic extinction coefficient, c(H$\beta$), is usually derived using the H$\alpha$/H$\beta$ ratio for case B recombination and assuming standard values of electron density and temperature. However, the use of strong Balmer lines can lead to selection biases when studying regions with different surface brightness, such as extended nebulae, with the use of single integral field spectroscopy observations, since, in some cases, the H$\alpha$ line can be saturated in moderate to long exposures. In this work, we present a method to derive extinction corrections based only on the weaker lines of HeI, taking into account the presence of triplet states in these atoms and its influence on recombination lines. We have applied this procedure to calculate the extinction of different regions of the 30 Doradus nebula from MUSE integral-field spectroscopy data. The comparison between helium and hydrogen c(H$\beta$) determinations has been found to yield results fully compatible within the errors and the use of both sets of lines simultaneously reduces considerably the error in the derivation.
\end{abstract}

 
\section{Introduction} 
\label{sec:intro}

The study of the chemical abundances of giant extragalactic HII regions (GEHR) provides most of our knowledge of these abundances in external galaxies. However their accurate determination requires the detection and measurement of temperature dependent weak emission lines. If that is not possible, different calibrations of mainly oxygen abundances have been proposed, none of them entirely satisfactory and producing abundance derivations with errors of 0.2-0.3dex (a recent summary of this can be found in \citet{2022MNRAS.511.4377D} and a comparison of the different calibrations used is given in \citet{2005MNRAS.361.1063P}). This precludes any spatial study of chemical inhomogeneity across HII regions.

During many years observations of GEHR regions have been obtained with the use of long slit spectroscopy and have been focused only in the brightest regions, hence integrated spectra have been weighted by luminosity and/or surface brightness. Thus, up to now, HII regions have been usually typified by only their brightest regions although it is widely recognised that the underlying assumption that the physical conditions and abundances of the brightest part of a nebula are representative of the whole entity, still remains questionable. In fact, low surface brightness regions are rarely observed and there are few studies of spatially resolved HII regions from long slit observations (but see, for example NGC 5471 in M101 by \citet{1985ApJ...290..449S};  NGC 604 in M33 by \citet{1987MNRAS.226...19D} and  \citet{2004AJ....128.1196M}; and 30 Doradus by \citet{1987ApJ...317..163R}). Now with the use of integral field spectroscopy (IFS) this can be done much more easily so that we can learn about variations of physical conditions throughout an entire nebula \citep{2010MNRAS.408.2234G,2013MNRAS.430..472L}, provided we can obtain all the elements to do it, for which an essential step is the interstellar extinction correction of the observed emission line fluxes from which the gaseous physical conditions, electron density and temperature, and the elemental abundances are derived.

Interstellar extinction is generated by the scatter and absorption of photons by the medium between the observer and the radiation source. It is produced by dust grains and gas present in the interstellar medium and it is proportional to the gas column in the line of sight. This mechanism, Mie scattering, is effective in photons with wavelengths comparable to the size of grains and is less efficient in photons of longer wavelengths. Thus, the effect is dependent on wavelength and the radiation of the source appears reddened.


The correction of the observed line fluxes for this effect is an indispensable preliminary step for the physical interpretation of the data. The total extinction at the corresponding wavelength can be calculated by comparing the flux of lines with the continuum radio-frequency emission. Another correction method is to compare known intensity ratios of different emission lines which have a low dependence on physical conditions, essentially density and temperature. This procedure provides a differential extinction between the lines used. Through the years, different methods to determine relative line extinction affecting observations have been proposed: 

(i) Calculating the ratio of two forbidden lines from the same upper set of levels \citep[e.g. sulphur two forbidden line,][]{1968ApJ...154L..57M}. This procedure has two main problems: the involved lines are moderately weak and the wavelength range is very wide, hence it is not usually covered by a simultaneous observation.

(ii) Comparing hydrogen recombination lines from two different series, for example, Balmer and Paschen, arising from the same upper level. Under this condition, the line intensity ratios depend mainly on their transition probabilities, i.e. the Einstein coefficients, and the gas physical conditions are not relevant. This method is excellent because the involved line ratios are almost constant for the typical range of nebular temperatures and densities observed in HII regions \citep[see Tab. 4.4 of][]{Osterbrock2006}, the most important disadvantage being the large difference between the series intensities, which implies additional errors in the extinction determinations. 

(iii) Comparing two hydrogen recombination lines from the same series, even if they do not originate from same upper level. This is the case of H$\alpha$/H$\beta$, H$\beta$/H$\gamma$, etc. The lines are present in all regions of nebulae and they are sufficiently strong. This method is the most widely used to correct emission line intensities for interstellar absorption. Some assumptions about the recombination theory scheme are required, as well as the previous knowledge of the theoretical line ratios employed. Although these ratios depend on the density and electron temperature of the nebula, this dependence is rather weak \citep[see Tab. 4.2 of][]{Osterbrock2006} and therefore the method constitutes an acceptable approach.

However, due to the fact that some of the already existing spectrographs and others coming up in the future, encompass a wavelength range not including hydrogen Balmer lines other than H$\beta$ and H$\alpha$, the reddening constant determination relies only on the ratio between these two lines with the former one being, in many cases, heavily affected by underlying stellar absorption, and thus producing a very uncertain extinction correction. 
In this case adding the HeI recombination lines to the analysis can provide a much more reliable determination.

Also in the case of extended nebulae IFS provides images showing a very large dynamical range of surface brightness hence, in many cases, it is not possible to analyse the region in its full extent using a single exposure: a shorter one has a good S/N in the strong nebular emission lines, including both H$\alpha$ and H$\beta$, but the weak emission lines, as for example the auroral electron temperature sensitive lines crucially needed for the determination of the physical conditions of the gas, can be measured with a reasonable S/N only for the brightest filaments. On the other hand, in longer exposures, these weak lines can be detected and measured easily, but the strongest lines, including H$\alpha$, are often saturated, not allowing the determination of the extinction if only the hydrogen lines are used. In this second case, we propose the use of HeI lines to derive the reddening constan despite the fact that the HeI lines are weaker, since it circumvents saturation effects.

In this work we present a method to derive the logarithmic extinction coefficient at optical wavelengths using the emission lines of HeI checking for consistency between the proposed method and the commonly used procedure of using HI Balmer recombination lines and showing that, in fact, both sets of lines can be used jointly providing a reliable determination. Section~\ref{theoretical context}  addresses the theoretical aspects involved in the methodologyand is included for the benefit of the reader not familiar with the atomic physics involved. Section~\ref{data} provides an application of the method based on observations of the central part of the very well studied GEHR 30Dor in the Large Magellanic Cloud (LMC) obtained with the MUSE spectrograph attached to the VLT telescope. Our results are presented and discussed in Section~\ref{results} and finally the summary and conclusions are presented in Section~\ref{summary}.

\section{Theoretical analysis}\label{theoretical context}
\subsection{Theoretical context}
\begin{figure}
\begin{center}
\hspace{0.25cm}
\includegraphics[width=\columnwidth]{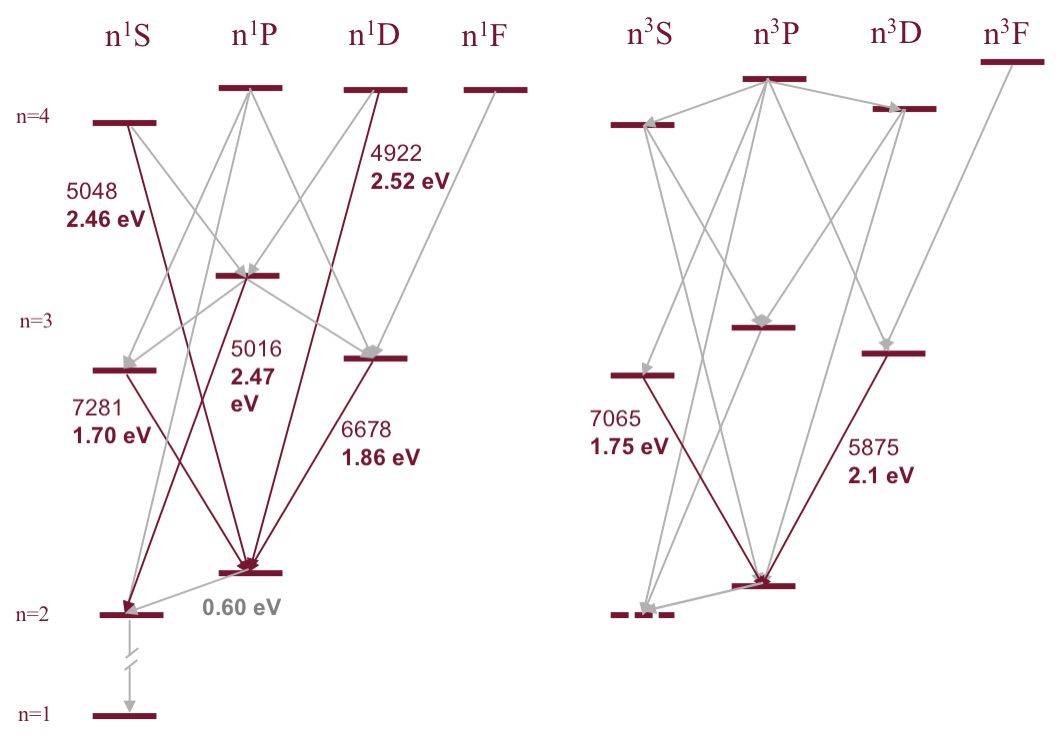}
\caption{Levels' structure of HeI. The left diagram is to singlet states and the right is to triplet stages. We have highlighted lines that we have analysed in this study.}
\label{fig:1}
\end{center}
\end{figure}

Fig. \ref{fig:1} shows the energy level structure of the HeI atom. It is similar to that of hydrogen except that
there are no sub-level degeneracies with different angular momentum for HeI. 

The He atom is a two-electron system and it has singlet and triplet levels. For singlet states, radiative recombination transitions cascade down to the 2$^1$P and 2$^1$S levels. These are the characteristics of the produced transitions: (i) The 2$^1$P level involves $\lambda \lambda$4922, 5048, 6678 and 7281 \AA\ lines. Its deexcitation can occur by two mechanisms: the emission of an infrared photon (2$^1$P$\rightarrow$2$^1$S transition) or the emission of a L$\alpha$ photon ( 2$^1$P$\rightarrow$1$^1$S transition) with their relative probabilities being given by the ratio of their Einstein coefficients, A($2^1P,\ 1^1S$)/A($2^1P,\ 2^1S)$ $\sim$ 1000, i. e. for every 1000 Ly$\alpha$ photons, 1 infrared photon is generated. In addition, since the cross-section of hydrogen atoms at the energy of the HeI Ly$\alpha$ photon is large, (a$_{21.3eV}$(H) $\sim$ 1.5$\cdot$10$^{-18}$ cm$^{-2}$) its absorption by an HI atom is more probable than 1000 Ly$\alpha$ absorptions and re-emissions. We can therefore assume that this last one is the dominant process, it is optically thick and the case B recombination is an appropriate approximation. (ii) The  2$^1$S$\rightarrow$1$^1$S transition producing the $\lambda$5016 \AA\ line occurs through the emission of two continuum photons (20.7 eV) with the relative probability of one of them having an energy equal to or higher than 13.6 eV being 0.56. Therefore, it would be able to ionize H and contribute to the diffuse radiation field. 

For triplet states, radiative recombination cascades down to the 2$^3$S metastable level. The decay of these electrons to the fundamental level 1$^1$S is forbidden, with probability of A(2$^3$S, 2$^1$S) = 1.26$\cdot$10$^{-4}$s$^{-1}$ \citep{Osterbrock2006}. The 2$^3$S level can be collisionally depopulated to the singlet levels 2$^1$S and 2$^1$P, with a high probability (and with a spin change). The depopulation to the 1$^1$S level is also possible but with very low probability. The critical electron density for this process to occur, n$_c$(2$^3$S), is defined as the balance between the probability of radiative and collisional transitions:
\begin{equation}\label{eq:nc}
n_c(2^3S) =\frac{A(2^3S,\ 1^3S)}{C(2^3S,\ 2^1S)+C(2^3S,\ 2^1P)} 
\end{equation}
where A and C are the corresponding Einstein and collisional coefficients respectively. Values of C(2$^3$S, 2$^1$S) and C(2$^3$S, 2$^1$P) range from 1.95$\cdot$10$^{-8}$ cm$^{3}$s$^{-1}$ to 2.68$\cdot$10$^{-8}$ cm$^{3}$s$^{-1}$ and 2.34$\cdot$10$^{-9}$ cm$^{3}$s$^{-1}$ to 9.81$\cdot$10$^{-9}$ cm$^{3}$s$^{-1}$ respectively for temperatures between 6000 and 25000 K \citep[see Tab. 2.5 of ][]{Osterbrock2006}. Therefore the critical electron density ranges from 6.2 $\times$ 10$^3$ cm$^{-3}$ to 3.34 $\times$ 10$^3$ cm$^{-3}$. In HII regions, the electron density is usually of the order of 10$^2$cm$^{-3}$ and lower than n$_c$, thus radiative transitions dominate. 

However, even at lower densities, n$_e$ < n$_c$(2$^3$S), the collisional contribution can play a significant role for several lines. This is the case of the $\lambda$7065 \AA\ line for which the ratio of collisional to recombination contributions is 0.152 as compared to the value of 0.054 found for the $\lambda$5876 \AA\ line. For singlet lines, collisional corrections are 0.013, 0.015, <0.01, 0.016 and 0.054 for the $\lambda\lambda$ 4922, 5016, 5048, 6678 and 7281 \AA\ lines respectively \citep[][T = 20000K and n$_e$ = 100cm$^{-3}$]{Benjamin1999} calculated using the radiative cascade rates of \citet{1996MNRAS.278..683S} and the collisional rates of \citet{1993ADNDT..55...81S}.

Additionally, since 2$^3$S is a metastable level, the effect of radiative transfer can also be important in this study. A large population of electrons in this level can generate a significant optical depth, particularly in n$^3$P$\rightarrow$2$^3$S transitions, and the revision of the assumption of Case B recombination is necessary. For instance, members of the n$^3$P series can be transformed into members of other series (n$^3$S and n$^3$D). This effect increases with the number of electrons in the 2$^3$S level, which depends directly on the probability of their transition to levels 2$^1$S and 2$^1$P, requiring threshold energies of 0.9 eV and 1.5 eV respectively. We can define the optical depth factor to radiative transfer for a given line, f$_{line}$(T, n$_e$, $\tau$), as the ratio between the emissivity of the line for given values of T, n$_e$ and $\tau$ and the emissivity for the same T and n but with $\tau$ = 0. Even for densities as high as n$_e$ = 10$^{8}$ cm$^{-3}$, the optical depth factor on singlet lines is less than 0.4\% at worst and for the $\lambda$5876 \AA\ line  (T = 10000 K, n$_e$ = 100 cm$^{-3}$, $\tau$ = 100) takes an accepted value of 1\%. Nevertheless, the $\lambda$7065 \AA\ line shows a strong dependence on this effect, of almost 4\% \citep{Benjamin2002}. 
 
\subsection{Interstellar extinction correction using the ratio of two recombination lines}

Light is absorbed as it travels through the interstellar medium in such a way that the observed flux of each spectral emission line follows the equation: 
\begin{equation}\label{eq:1}
F _\lambda = I_\lambda \cdot 10^{-0.4\cdot A_\lambda}
\end{equation}
\noindent where F$_\lambda $ and I$_\lambda $ are the observed and emitted fluxes at wavelength $\lambda$ and A$_\lambda $ is the corresponding extinction in magnitudes along the observing path. Written as a function of the optical depth, $\tau _\lambda$:
\begin{equation}\label{eq:2}
F _\lambda = I_\lambda \cdot e^{-\tau _\lambda} 
\end{equation}
where $\tau _\lambda$ depends on wavelength and the physical properties of the medium, i. e. the column density of absorbing dust grains and the extinction cross-section per particle. This is expressed by means of  an extinction curve which gives $\tau _\lambda = C\cdot f(\lambda)$. 

The observed flux can be normalised to the flux of a reference line. For extinction corrections derived from HI lines, the reference line is usually H$\beta$ and therefore the value of the logarithmic extinction curve at this wavelength is null, f(H$\beta$) = 0. With this normalisation, equation \ref{eq:2} is written as:
\begin{equation}\label{eq:3}
\frac{F_\lambda}{F_{H\beta}}=\frac{I_\lambda}{I_{H\beta}}\cdot 10^{-0.434\cdot C(f(\lambda)-f(H\beta))}
\end{equation}
For extinction corrections derived from HeI lines we propose to use as reference line the HeI $\lambda$6678 \AA\ line, the most prominent of the singlet levels. In this case, for equation \ref{eq:2} we obtain:
\begin{equation}\label{eq:4}
\frac{F_\lambda}{F_{HeI\lambda 6678}}=\frac{I_\lambda}{I_{HeI\lambda 6678}}\cdot 10^{-0.434\cdot C(f(\lambda)-f(HeI\lambda 6678))}
\end{equation}

The logarithmic extinction coefficient, c(H$\beta$), is the reddening constant defined as c(H$\beta$) = 0.434$\cdot$ C. With this definition, equations \ref{eq:3} and \ref{eq:4} are given by:
\begin{equation}\label{eq:5}
\frac{F_\lambda}{F_{H\beta}}=\frac{I_\lambda}{I_{H\beta}}\cdot 10^{-c(H\beta)(f(\lambda)-f(H\beta))}
\end{equation}
\begin{equation}\label{eq:6}
\frac{F_\lambda}{F_{HeI\lambda 6678}}=\frac{I_\lambda}{I_{HeI\lambda 6678}}\cdot 10^{-c(H\beta)(f(\lambda)-f(HeI\lambda 6678))}
\end{equation}

The final logarithmic extinction coefficient can be calculated as:
\begin{equation}\label{eq:7}
c(H\beta) = -\frac{1}{f(\lambda)-f(H\beta)}\cdot \left[ log \left(\frac{F_\lambda}{F_{H\beta}}\right)- log \left(\frac{I_\lambda}{I_{H\beta}}\right) \right]
\end{equation}
for the hydrogen, with f(H$\beta$) = 0, and
\begin{equation}\label{eq:8}
\begin{split}
c(H\beta) = & -\frac{1}{f(\lambda)-f(HeI\lambda 6678)}\cdot \\ & \cdot \left[ log \left(\frac{F_\lambda}{F_{HeI\lambda 6678}}\right) -log \left( \frac{I_\lambda}{I_{HeI\lambda 6678}}\right) \right]
\end{split}
\end{equation}
for the helium correction method.

\begin{table}
\centering
\begin{tabular}{|c|c|c|}
\hline 
Line              & f($\lambda$)$^a$& I$_\lambda$/I$_{ref}$ $^b$\\ \hline
H$\beta$ & 0 & 1.000\\
HeI $\lambda$4922 \AA\ & -0.014 & 0.345\\
HeI $\lambda$5875 \AA\ & -0.209 & 3.518\\
H$\alpha$ & -0.313 & 2.863\\
HeI $\lambda$6678 \AA\ & -0.329 & 1.000\\  
HeI $\lambda$7065 \AA\ & -0.377 & 0.621\\
HeI $\lambda$7281 \AA\ & -0.402 & 0.187\\ 
Pa17 & -0.514 & 0.004\\
Pa16 & -0.518 & 0.004\\
Pa15 & -0.521 & 0.005\\
Pa14 & -0.525 & 0.007\\
Pa13 & -0.531 & 0.008\\
Pa12 & -0.537 & 0.011\\
Pa11 & -0.546 & 0.014\\
Pa10 & -0.557 & 0.018\\
Pa9 & -0.572 & 0.025\\
 \hline 
\end{tabular}
\begin{tablenotes}
\centering
\item $^a$ \cite{1972ApJ...172..593M}, R$_v$ = 3.2 .
\item $^b$ \citet{pyneb}, n$_e$(cm$^{-3}$) = 10$^2$, T(K) = 10$^4$.
\end{tablenotes} 
\caption{Empirical values of the normalised logarithmic extinction at the wavelengths of the different HeI emission lines from \citet{1972ApJ...172..593M}, and the ratio of these lines to the  $\lambda$ 6678 \AA\ reference line as calculated by the PyNeb \citep{pyneb} tool for T= 10$^4$ K and n$_e$= 10$^2$ cm$^{-3}$, using \citet{Storey1995} and \citet{2012MNRAS.425L..28P} atomic data for hydrogen and helium respectively.}
\label{Tab:1}
\end{table}

The values of f($\lambda$) and the theoretical ratios between the different helium and hydrogen lines and the reference ones at $\lambda$ 6678 \AA\ and H$\beta$ respectively are given in Tab. \ref{Tab:1}.

\subsection{Interstellar extinction correction using a linear regression}

Equations \ref{eq:5} and \ref{eq:6} can be edited as:
\begin{equation}\label{eq:9}
log_{10}\left(\frac{F_\lambda }{F_{H\beta }}\right)- log_{10}\left(\frac{I_\lambda }{I_{H\beta }}\right)= -c(H\beta )(f(\lambda )-f(H\beta ))
\end{equation}

\begin{equation}\label{eq:10}
\begin{split}
log_{10}\left(\frac{F_\lambda}{F_{HeI\lambda6678}}\right)& - log_{10}\left(\frac{I_\lambda}{I_{HeI\lambda6678}}\right)
 = \\  & = -c(H\beta)(f(\lambda) -f(HeI\lambda6678))
\end{split}
\end{equation}
These equations are equivalent for the same reddening curve normalisation (in this case, f(H$\beta$) = 0). The difference between the observed and theoretical logarithmic ratios constitutes the dependent variable, the normalised logarithmic extinction at each wavelength represents the independent variable and the logarithmic extinction coefficient, c(H$\beta$) is the slope of the fitting.

\section{Observations and measurements}\label{data}

\begin{table*}
\begin{tabular}{|c|c|c|c|c|c|c|}
\hline
Field ID & \begin{tabular}[c]{@{}c@{}}RA \\ (hh:mm:ss)\end{tabular} & \begin{tabular}[c]{@{}c@{}}Dec \\ (dd:mm:ss)\end{tabular} & Exp. time (s)&
ESO project ID & \begin{tabular}[c]{@{}c@{}}Observation date \\ (yy-mm-dd, hh:mm:ss)\end{tabular} & Seeing (”)\\ \hline

30Dor-A & 05:38:36.75 & -69:06:32.8 & 240 & 60.A-9351 &  2014-08-19, 08:54:50 & 1.240 \\ 

30Dor-A & 05:38:36.73 & -69:06:32.8 & 2400 & 60.A-9351 &  2014-08-19, 09:10:05 & 1.262\\ 

30Dor-B & 05:38:47.37 & -69:06:32.8 & 240 & 60.A-9351 &  2014-08-22, 08:57:44 & 1.507 \\ 

30Dor-B & 05:38:47.36 & -69:06:32.8 & 2400 & 60.A-9351 &  2014-08-22, 09:07:05 & 1.343 \\ 

30Dor-C & 05:38:36.67 & -69:05:34.8 & 240 & 60.A-9351 &  2014-08-24, 08:46:09 & 1.499\\ 

30Dor-C & 05:38:36.66 & -69:05:34.8 & 2400 & 60.A-9351 &  2014-08-24, 08:55:31 & 1.415\\ 

30Dor-D & 05:38:47.44 & -69:05:34.8 & 240 & 60.A-9351 &  2014-08-25, 08:55:04 & 1.252\\ 

30Dor-D & 05:38:47.43 & -69:05:34.7 & 2400 & 60.A-9351 &  2014-08-25, 09:04:20 & 1.188\\ \hline

\end{tabular}
\caption{Description of the observations.}
\label{tab:0}
\end{table*}
We have used observations of the central part of 30 Dor obtained with the MUSE spectrograph \citep[Multi-Unit Spectroscopic Explorer][Program ID: 60.A-93]{MUSE} to test the proposed method. Four fields have been analysed, each of them having exposure times of 240 s and 2400 s, a Field of View (FoV) close to 1.52' and a pixel scale of 0.2". The observations were made on the 19th, 22nd, 24th, and 25th of August 2014 (see Tab. \ref{tab:0}). 

The IFU observations of this nebula provide simultaneously areas with very different surface brightness. While in long slit observations high surface brightness regions are preferentially observed and aperture effects are important, the use of IFU data can circumvent these disadvantages. Additionally, the extremely high quality of the chosen two sets of MUSE observations makes them ideal for the purpose of comparing the two methods described here for extinction derivation. The high S/N obtained in the weakest lines from the long exposure cubes is comparable to the typical S/N of data employed for the study of extinction in emission nebulae using strong recombination lines. Also, the high spectral resolution of MUSE data (R $\sim$ 1770 at 4800 \AA\ and R $\sim$ 3590 at 9300) contributes to this critical aspect increasing the S/N ratio linearly with the resolution.

We have made the region selection measuring the fluxes of strong and weak emission lines in spectra extracted from the short (240 s) and long (2400 s) exposure cubes respectively. These two approaches are complementary since in the short exposure spectra the faint emission lines are almost undetectable in low surface brightness regions, while strong lines are saturated in the long exposure spectra of high surface brightness ones. Thus, H$\alpha$ and H$\beta$ emission lines have been measured from cubes with 240s of exposure time while HeI and Paschen HI lines have been measured from 2400 s exposure cubes. Random spectra have been extracted simulating a circular aperture of 1.5" diameter that corresponds to the largest seeing value of the observations (see \ref{tab:0}). The area covered by the selected data amounts to more than 35\% of the total area covered by the 4 data cubes. To guarantee the star formation origin of the extracted spectra we have imposed the requirements:  EW(H$\alpha$) >  6 \AA\ \citep[][]{CidFernandez2010, Sanchez2015} and 2.7 < H$\alpha$/H$\beta$ < 6 \citep[][n$_e$ = 100 cm$^{-3}$, T$_e$ = 10$^4$ K]{Osterbrock2006}.

\begin{figure*}
\begin{center}
\includegraphics[width=0.9\textwidth]{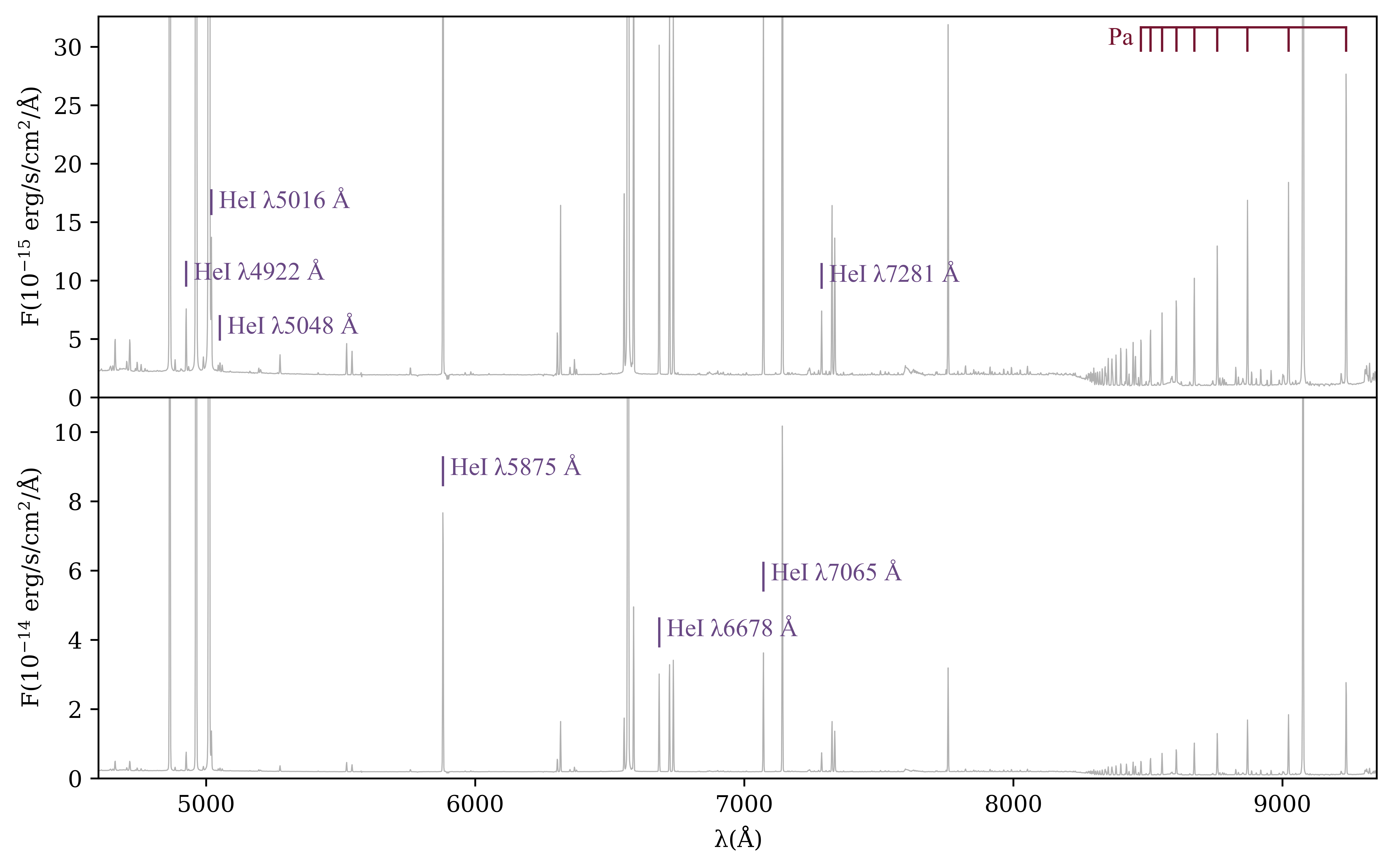}
\caption{HeI and Paschen emission lines in the optical spectrum from data in the 2400 s exposure. Upper panel is a zoom of the bottom panel to see weak lines.}
\label{fig:3}
\end{center}
\end{figure*}

Fig.\ref{fig:3} shows a typical spectrum extracted from the analysed cubes, marking the  HeI emission lines present in the wavelength range covered by the data (from 4800 to 9300 \AA). Three lines have been discarded for the analysis: $\lambda$ 5016 \AA\ overlaps with [OIII]$\lambda$ 5007 \AA, $\lambda$ 5048 \AA\ is too weak, and $\lambda$ 7065 \AA\ suffers from optical depth  and collisional excitation effects. Finally, we have used the lines $\lambda$4922, $\lambda$5876, $\lambda$6678, $\lambda$7281 \AA\ of HeI, and H$\alpha$, H$\beta$ and Paschen lines of HI, these latter ones for comparison purposes.

The line flux measurements have been performed as described next: (i) the emission lines present in the spectrum have been masked with a width of $\pm$8 \AA\ around the central wavelength of the line; (ii)  a global continuum has been adjusted fitting a second-order polynomial and the statistical dispersion of the continuum, $\sigma_c$, has been calculated; (iii) we have assigned a local continuum to each emission line to be measured, taking into account the dispersion of the global continuum fitting, $F_c(line)=F_c(\lambda) \pm  \sigma_c$; (iv) finally, we have fitted a Gaussian function according to the local continuum level. 
The stellar absorption underlying the H$\alpha$, H$\beta$ and HeI lines has been corrected using, in step (iv), a two-component Gaussian fit to reproduce simultaneously emission and absorption in each line \citep[see ][]{Diaz2007}.

Flux errors have been calculated from the expression given in \citet{Gonzalez-Delgado1994}:
\begin{equation}\label{eq:err}
\Delta [F_\lambda] =\sigma _l \cdot N^{1/2}[1+EW/(N \Delta)]^{1/2}
\end{equation}
where $\Delta [F]$ is the error in the line flux, $\sigma _l$ represents the standard deviation in the continuum near each line, N is the number of pixels used in the measurement of the line flux, EW is the line equivalent width and $\Delta$ is the spectral dispersion (1.25 \r{A}/pix). This expression takes into account the photon statistics and the error in the continuum of each line. The calculated errors have been propagated in quadrature for the rest of the derived quantities.

We have assumed the \cite{1972ApJ...172..593M} law with a specific attenuation of R$_v$ = 3.2, case B of recombination and a simple screen distribution of dust. The ratio of helium and hydrogen lines have been selected from the Tab. \ref{Tab:1} and references explained in the table.

\section{Results and discussion}
\label{results}

\begin{table}
\centering
\caption{Extraction parameters for emission line maps.}
\label{tab:line ranges}

\begin{tabular}{lcccc}
\hline
Line &  $\lambda_c$ (\r{A})  & $\Delta\lambda$ (\r{A}) & $\Delta\lambda_{left}$ (\r{A}) & $\Delta\lambda_{right}$ (\r{A}) \\ 
\hline
 H$\alpha$ & 6563 & 8& 6523.0 - 6542.0 & 6596.0 - 6612.0 \\
 H$\beta$  & 4861 & 8& 4827.5 - 4843.5 & 4892.4 - 4908.4 \\
 $[SIII]$ & 6312 & 8& 6236.2 - 6252.2 & 6406.0 - 6422.0 \\
 HeI & 6678& 8 & 6645.8 - 6661.8 & 6681.8 - 6697.8\\
\hline
\end{tabular}
\begin{tablenotes}
\item All wavelengths are in rest frame.
\end{tablenotes}
\end{table}

We have constructed 2D maps from the observed data cubes. For the different emission lines we have assumed a linear behavior of the continuum emission in the region of interest choosing side-bands around each line of a given width. Table \ref{tab:line ranges} give the identification of each line in column 1, its central wavelength, $\lambda_c$ in \AA\ in column 2, its width, $\Delta\lambda$, in \AA\ in column 3, and the wavelength range of the two continuum side-bands, in \AA\ in columns 4 and 5. We have not taken into account the underlying stellar absorption in HeI and HI lines since, in theis case,their contribution is negligible.

\begin{figure*}
\begin{center}
\includegraphics[width=0.75\textwidth]{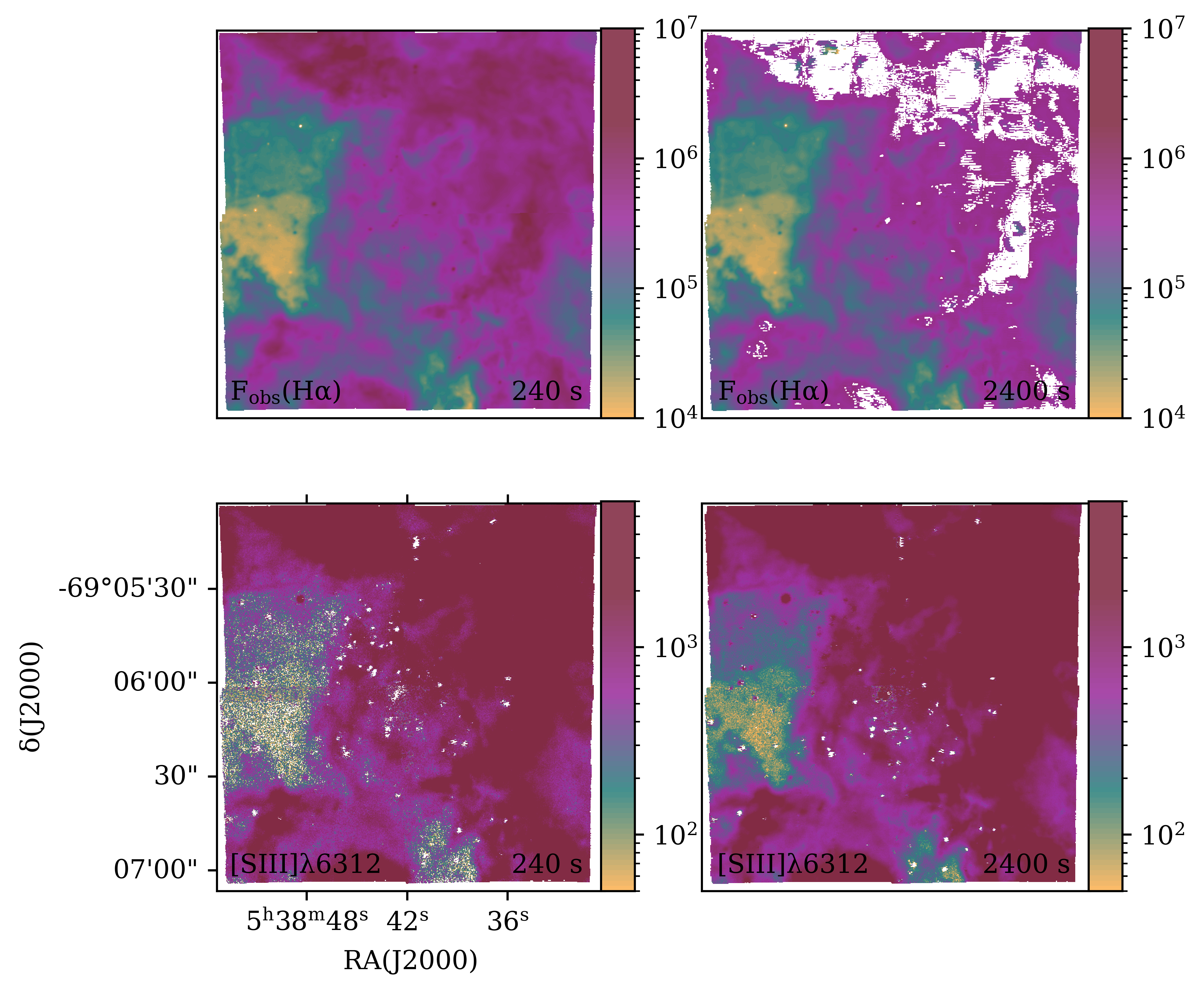}
\caption{H$\alpha$ (upper panels) and [SIII] $\lambda$ 6312 \AA\ (lower panels) line maps obtained from short (left) and long (right) exposure MUSE data. Fluxes are given in units of $10^{-20}$ $erg/s/cm^2$. The white areas in the upper right panel show the saturation effects in the H$\alpha$ line in the long exposure frame. The lower left panel shows the practically non detection of the [SIII] $\lambda$ 6312 \AA\ (at the noise level) in the low surface brightness region  in the short exposure one.}
\label{fig:3b}
\end{center}
\end{figure*}

Fig. \ref{fig:3b} shows the selection effect due to the large dynamical range of luminosity and surface brightness of GEHR in the two sets of constructed maps. The images in the upper panels correspond to short (left) and long (right) exposure images in the H$\alpha$ lines. The white colour indicates the areas where the H$\alpha$ line is completely saturated and thus the value of the extinction cannot be obtained, hence the emission line ratios cannot be corrected for reddening. One might think the left image is good enough to allow a good nebular analysis. However, the images in the lower panel, that correspond to the weak auroral line of  [SIII] at $\lambda$ 6312 \AA , show a S/N so low in the short exposure one (left) that, in many regions, the line cannot even be detected, and so only the brighter regions can be fully analysed. On the other hand, in the long exposure image (right) the S/N is more than reasonable all across the frame showing that a complete characterization of the gas can be made using only relatively weak emission lines if extinction can be calculated without the use of the H$\alpha$ line, that is using HeI lines instead.

\begin{figure*}
\begin{center}
\includegraphics[width=\textwidth]{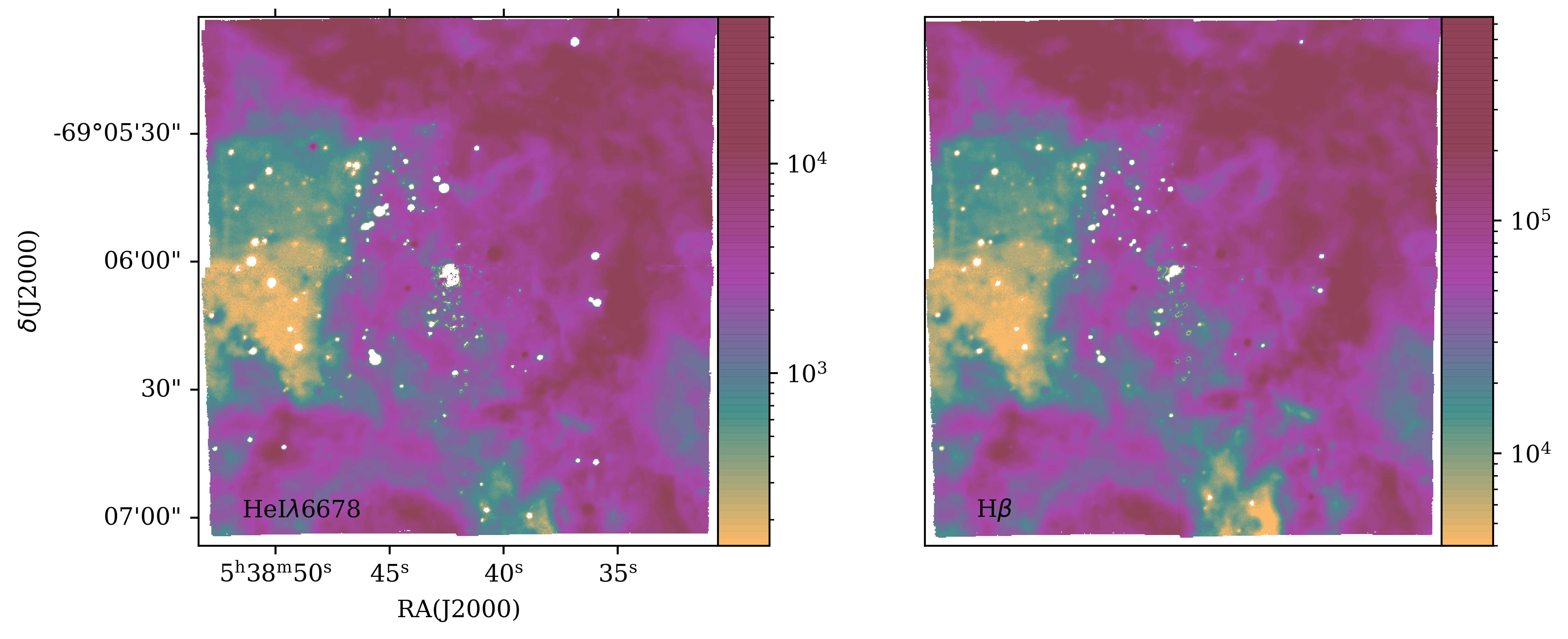}
\caption{Reference line maps for hydrogen and helium. Left: observed HeI $\lambda$6678 \AA\ flux. Right: observed H$\beta $ flux. All images are in logarithmic scale. North is up and East is to the left. Fluxes are given in units of $10^{-20}$ $erg/s/cm^2$.}
\label{fig:6-7}
\end{center}
\end{figure*}

\begin{figure*}
\begin{center}
\hspace{0.25cm}
\includegraphics[width=0.65\textwidth]{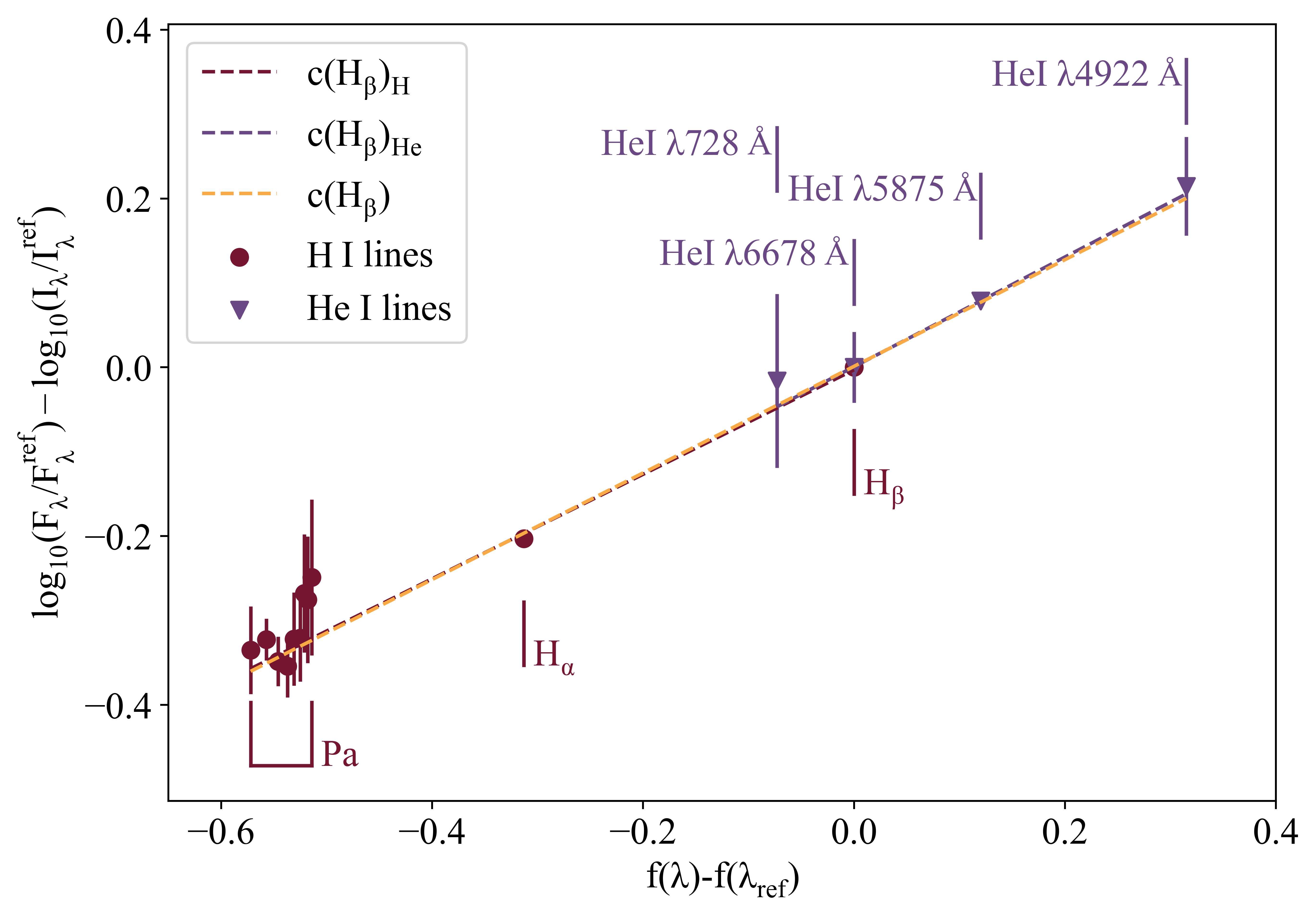}
\caption{Linear regression of c(H$\beta$) values from hydrogen lines, helium lines and both sets of lines together. H$\alpha$ and HeI $\lambda$5875 \AA\ errors are inside the symbols in the graph.}
\label{fig:4}
\end{center}
\end{figure*}

We have checked the spatial distribution of the H$\beta$ and HeI $\lambda$ 6678 \AA\ lines used for normalisation in the extinction derivation method (see Fig. \ref{fig:6-7}). Both maps show the same spatial distribution across the nebula and very similar S/N. The pixel-by-pixel ratio of H$\beta$ to HeI $\lambda $6678 \AA\ calculated from the maps has a mean value of $\sim$ 20 with a standard deviation of 5 which corresponds to  a 0.113\% and a 0.006\% of the mean values of the HeI $\lambda $6678 \AA\ and H$\beta$ intensity maps respectively.

For short exposure spectra (240 s), we have used the ratio of the H$\alpha$ and H$\beta$ lines to derive the mean extinction (Eq. \ref{eq:7}). The estimated mean errors in the line fluxes are 0.92\% for H$\beta$ and 0.15\% for H$\alpha$, producing values of c(H$\beta$) with an absolute mean error of 0.02. In the case of the helium lines, the long exposure data (2400 s) have been used choosing the two strongest HeI lines: $\lambda \lambda $ 5875 and 6678 \AA\ with associated mean errors of 0.37\% and 1.37\% respectively which yield values of c(H$\beta$) with an absolute mean error of 0.05 (Eq. \ref{eq:8}).  

\begin{table}
\centering
\begin{tabular}{|c|c|}
\hline 
Fit & c(H$\beta$) \\ \hline
Hydrogen & 0.619 $\pm$ 0.018 \\
Helium & 0.650 $\pm$ 0.050 \\
Hydrogen  and Helium & 0.632 $\pm$ 0.010\\
 \hline 
\end{tabular}
\caption{Reddening constant values obtained from the fitting shown in Fig. \ref{fig:4} for a randomly selected emission region.} 
\label{tab:2}
\end{table}

In order to minimise errors in the reddening constant determination, we have fitted a linear regression using all the available H and HeI lines in Eqs. \ref{eq:9} and \ref{eq:10}, following  \citet{Diaz2007} for hydrogen. Fig. \ref{fig:4} shows an example of the application of this method to the spectra of one selected emission region extracted from two cubes of different exposure times (240 s and 2400 s). The fits obtained for hydrogen lines, helium lines and all the lines together are consistent with each other and the intercept is compatible with zero, with the fit using all the lines yielding the smaller error for the c(H$\beta$) determination (see Tab. \ref{tab:2}). Although the error found using only the HeI lines is larger than that using the HI lines, it should be taken into account that the underlying stellar absorption is less important than in the case of hydrogen analysis \citep[see][]{2005MNRAS.357..945G,2021JCAP...03..027A}. If the underlying absorption were a critical issue in our work, a simultaneous fit of decrements and stellar absorption could be done \citep[see, for example][]{1994ApJ...435..647I}.

Also, in order to check the reliability of our results, we have introduced the derived methodology in the Bayesian algorithm described in \citet{2019MNRAS.487.3221F} which simultaneously fits a 14 parameters chemical model, obtaining results which are fully compatible with those reported here.

Finally, we have run a bootstrap model using random temperatures and densities in order to evaluate the effect of these physical parameters on the HeI lines. The bootstrap method is a statistic  procedure that can be used to estimate characteristics of a calculated parameter, like the mean or standard deviation, resampling with replacement the involved variables. In this work we have used 10$^6$ repeats and we have assumed a Gaussian distribution of density with $\mu$ = 275.5 cm$^{-3}$  and $\sigma$  = 75.16 cm$^{-3}$, that correspond to densities in a range from 50 to 500 cm$^{-3}$ considering a 3$\sigma$ width. In the case of temperatures, we have assumed a Gaussian distribution with $\mu$ = 11000 K and $\sigma$ = 1000 K, that correspond to a temperature range from 8000 K and 14000 K. We have computed the ratio of H$\alpha$/H$\beta$ and $\lambda$5876 \AA\ /$\lambda$6678 \AA\ lines using PyNeb \citep{pyneb} and atomic data from \citet{Storey1995} and \citet{2012MNRAS.425L..28P} for hydrogen and helium respectively. The standard deviations for hydrogen and helium results are 0.0185 and 0.0041 respectively and the lines ratio distributions seem to be approximately gaussians. From these results, we can conclude that the ratio of HeI lines are less affected by temperature and density variations than the H$\alpha$/H$\beta$ ratio which represents another important advantage of the proposed method.

On the other hand, as mentioned in section \ref{theoretical context} there are some effects that need to be taken into account, in particular  collisional de-excitation, radiative transfer and optical depth of some levels. Regarding the first, the collisional contribution, before discarding the $\lambda$7065 \AA\ line, is within measurement errors and their effect is minimised in the fitting procedure.  This is also the case for the other two considered effects. 

\section{Summary and conclusions}
\label{summary}

In this article we have presented a method to derive the logarithmic extinction coefficient at optical wavelengths using the emission lines of HeI. The aim of this work has been two-folding. Firstly, due to the fact that some of the already existing spectrographs and others coming up in the future, encompass a wavelength range not including hydrogen Balmer lines other than H$\beta$ and H$\alpha$, the reddening determination relies only on the ratio between these two lines with the first of them being, in many cases, heavily affected by underlying stellar absorption. In this case we propose adding the HeI lines to the analysis in order to obtain a more reliable determination. Secondly, IFS provides images of extended gaseous nebulae with a very large dynamical range of surface brightness hence,in many cases, it is not possible to analyse the region in its full extent using a single exposure since a short one has a good S/N in strong nebular emission lines, including both H$\alpha$ and H$\beta$, but the weak emission lines needed for the determination of the physical conditions of the gas can be measured only for the brightest regions. On the other hand, in long exposures, these weak lines can be detected and measured easily, thus providing a very complete characterisation of the gas, but the strongest lines,including H$\alpha$, are saturated thus precluding the determination of the extinction if only HI lines are used. In this case we propose the use of HeI lines to obtain a good derivation of the reddening constant.

The method has been tested on archive MUSE data of the 30 Doradus nebula obtained with both short (240s) and long (2400s) exposure times. The reddening curve has been normalised to HeI $\lambda$ 6678 \AA\ and the calculations have been done assuming Case B recombination, a simple screen distribution of dust and the \citet{1972ApJ...172..593M} reddening law. We have compare the extinction derived for randomly selected regions using the weaker lines at $\lambda \lambda$ 4922, 5876, 6678, 7065 and 7281 \AA\ from the long exposure data and the ratio of the strong lines H$\alpha$ over H$\beta$ from the short exposure ones.

The comparison of the logarithmic extinction at H$\beta$ obtained using HI and HeI lines independently has proved both methods to be fully compatible within the errors although, as expected, these are slightly larger when using the weaker HeI lines. The errors in c(H$\beta$) get considerably reduced when both sets of lines are jointly used. The proposed methodology has been easily introduced into a Bayesian algorithm yielding results which are also compatible with those obtained by traditional computations.The fact that HeI and HI agree in the case of giant HII regions shows that none of the complexities related to line optical depth effects, departures from Case B recombination, or collisional excitation that might exist are important and the situation is classical and simple.

The results presented here will allow to study spatially resolved nebulae in their full extent, analysing both high and low surface brightness regions simultaneously in a single moderate-to-long exposure pointing.

\section*{Acknowledgements}

Based on data products from observations made with ESO Telescopes at the La Silla Paranal Observatory under programme ID 60.A-93. This research has been supported by the former Spanish Ministry of Economy, Industry and Competitiveness through the MINECO-FEDER research grant AYA2016-79724-C4-1-P, and the present Spanish Ministry of Science and Innovation through research grant PID2019-107408GB-C42. S.Z. acknowledges financial support from contract BES-2017-080509 associated to grant AYA2016-79724-C4-1-P. V. F. acknowledges support from FONDECYT Postdoc 2020 project 3200473 from the National Agency for Research and Development (ANID). We would like to thank Roberto Terlevich for the careful reading of this manuscript and his improvement suggestions and also an anonymous referee who, with his/her comments, has helped to clarify the aims of this work.

\section*{Data Availability}
The original data on which this article is based can be found in the ESO Science Archive Facility from ESO telescopes at La Silla Paranal Observatory.

\bibliographystyle{mnras}
\bibliography{bibliografia} 

\end{document}